\documentclass[12pt]{article}
\usepackage{heck}

\usepackage[utf8]{inputenc}
\usepackage{mathrsfs}
\usepackage{graphics,epsfig,color, graphicx}
\usepackage{verbatim}
\usepackage{CJK}
\usepackage{graphicx,epsf,amssymb,bbm,amsbsy,amsfonts,amssymb,amsmath}
\usepackage{hyperref}

\newcommand{\CPN}{\mathbb{CP}^{N-1}}
\newcommand{\SU}{\textrm{SU}}
\newcommand{\T}{\mathsf{T}}
\newcommand{\ii}{\mathsf{i}}

\begin{document}

%

%
%
%
%

\begin{CJK*}{UTF8}{min}

\title{Relating 't Hooft Anomalies of \\ 4d Pure Yang-Mills and 2d $\mathbb{CP}^{N-1}$ Model}

\preprint{IPMU-17-0157}

\authors{\centerline{Masahito Yamazaki (山崎雅人)}}

\institution{IPMU}{\centerline{Kavli IPMU (WPI), University of Tokyo, Kashiwa, Chiba 277-8583, Japan}}

\abstract{
It has recently been shown that a center-twisted compactification of the four-dimensional 
pure $\SU(N)$ Yang-Mills 
theory on a three-torus gives rise to the two-dimensional $\mathbb{CP}^{N-1}$-model on a circle
with a flavor-twisted boundary condition. We verify the consistency of this statement 
non-perturbatively at theta angle $\theta=\pi$, 
in terms of the mixed 't Hooft anomalies for flavor symmetries and the time-reversal symmetry. 
This provides further support for the approach to 
the confinement of four-dimensional Yang-Mills theory from the two-dimensional $\mathbb{CP}^{N-1}$-model.
}


\maketitle
\end{CJK*}

\section{Introduction}

The two-dimensional sigma model with the target space $\mathbb{CP}^{N-1}$
(the $\CPN$-model \cite{CPN})
has long been studied as a toy model for the four-dimensional Yang-Mills theory;
both theories are asymptotically free and has a dynamically-generated mass gap, for example.

Despite the similarities between the two, it has been long unclear
if there could be more direct and quantitative relations between the two.
Is the $\CPN$-model only a toy model?
Or could we use the $\mathbb{CP}^{N-1}$-model to actually solve the four-dimensional
Yang-Mills theory, e.g.\ for an analytical demonstration of confinement of the latter theory?\footnote{The relation between 
the two theories has been discussed in supersymmetric contexts, see  e.g.\ \cite{Cecotti,Dorey}.
Our focus here, however, is to study an honest non-supersymmetric theory where there is no protection from 
supersymmetry.}

A positive result in this direction has been recently given in the recent work of the author and 
K.\ Yonekura \cite{Yamazaki:2017ulc}. As we will explain below, it was shown
that a center-twisted compactification of four-dimensional pure $\SU(N)$ Yang-Mills theory 
on a three-torus $T^3= S^1\times T^2$, when the size of the two-torus $T^2$ is small,
gives rise to the two-dimensional $\mathbb{CP}^{N-1}$-model\footnote{A care is needed since this $\mathbb{CP}^{N-1}$ 
does not have the standard Fubini-Study metric.} on 
the residual circle $S^1$, with a flavor-twisted boundary condition studied in \cite{Dunne:2012ae}.
As emphasized in \cite{Yamazaki:2017ulc}  this provides a well-defined 
weakly-coupled setup with no notorious IR problems (e.g.\ the Linde problem \cite{Linde:1980ts}). 
This means that we in principle have a hope of
analytically-continuing back to the flat space $\mathbb{R}^4$---we first 
start in the weakly-coupled region and then 
sum up the perturbative as well non-perturbative contributions into a well-defined function 
(e.g.\ with the help of the Borel-\'Ecalle resummation of the trans-series \cite{Ecalle}, as applied to an infinite-dimensional setup),
and then adiabatically/analytically continue back to the flat $\mathbb{R}^4$ by smoothly changing the 
size of the torus.\footnote{This strategy of adiabatic continuation from compactification has a rather long history. The literature is too vast to be covered in this short note, see \cite{Adiabatic} for early references in eighties and see \cite{Yamazaki:2017ulc} for more references. Note that the novelty of the setup of \cite{Yamazaki:2017ulc} is to consider $T^2\times S^1$ compactification where there is a hierarchy between the sizes of $T^2$ and $S^1$.}

It would be, however, rather non-trivial to carry out this procedure in full generality,
and before venturing into the detailed computation
one might wish to further check the consistency of the result of \cite{Yamazaki:2017ulc},
preferably in a non-perturbative manner.

The goal of this short note is to verify the consistency of this proposal by matching the 't Hooft anomalies \cite{tHooft:1979rat} of the two theories
(in four and two dimensions). 
While the standard anomaly is lost upon dimensional reduction due to contamination from
high-energy states, we can study the recently-found  mixed 't Hooft anomaly 
between time reversal and center symmetry in four dimensions, present for the special value of the theta angle $\theta=\pi$ \cite{Gaiotto:2017yup}.
We find that this 't Hooft anomaly, upon a twisted compactification on a three-torus $T^3$,
precisely reproduces to the 't Hooft anomaly for a twisted compactification of the two-dimensional $\mathbb{CP}^{N-1}$ model on a circle,
as recently derived in \cite{Tanizaki:2017qhf}.

\section{From 4d to 2d}\label{sec.4d2d}
Let us first recapitulate the some of the crucial statements from \cite{Yamazaki:2017ulc}.
Consider four-dimensional pure $\SU(N)$ Yang-Mills theory 
on the geometry $\mathbb{R}\times T^3_{ABC}=\mathbb{R}\times S^1_A\times S^1_B\times S^1_C$,
with coordinates $t, x_A, x_B, x_C$. We denote the circumference of the circles by $L_{A}, L_B, L_C$,
and we choose the periodicity of $x_A, x_B, x_C$ to be $1$ ($x_A\sim x_A+1$ etc.).

In \cite{Yamazaki:2017ulc} the size of the circles are taken to be 
\begin{align}
L_A, L_B \ll L_C \;.
\label{hierarchy}
\end{align} 
In this parameter region the Yang-Mills gauge field along the two-torus $T^2_{AB}=S^1_A \times S^1_B$
is given by the flat connection on the two-torus, which is known to be parametrized by a 
point of the complex projective space $\mathbb{CP}^{N-1}$ \cite{FlatConnection}.\footnote{The complex projective space $\mathbb{CP}^{N-1}$ arising from the moduli space of flat connections do not have the canonical 
Fubini-Study metric, and in particular has singularities where new degrees of freedom of W-bosons emerge.
The analysis of \cite{Yamazaki:2017ulc}, however, 
was done at the classical values of the flat connection which are away from these singularities.
Such an analysis was enough to demonstrate the existence of fractional instantons
and the dynamical recovery of the center symmetry.
For this reason we expect that the subtlety coming from the singularities of the $\mathbb{CP}^{N-1}$ does not affect the 
discrete anomalies discussed in this paper.}
After reduction along the two-torus $T^2_{AB}$ we find that the resulting two-dimensional theory on $\mathbb{R}\times S^1_A$
is given by the two-dimensional $\mathbb{CP}^{N-1}$-model,
where the theta angle of the four-dimensional theory is identified with that of the 
two-dimensional theory. In this paper we denote the homogeneous coordinate of $\mathbb{CP}^{N-1}$
as $[z_1, \dots, z_N]$ (i.e.\ $[z_1, \dots, z_N]\sim [cz_1, \dots, cz_N]$ for $c\in \mathbb{C}^{\times}$).

Now the crucial ingredient of \cite{Yamazaki:2017ulc} is to include a $\mathbb{Z}_N$-center symmetry twist in the boundary conditions---we  include 't Hooft discrete magnetic flux \cite{tHooft:1979rtg} along the two-torus $T^2_{BC}=S^1_B\times S^1_C$.\footnote{Beaware that this is different from 
the other two-torus $T^2_{AB}=S^1_A\times S^1_B$.} It was then shown that this reproduces the twisted boundary condition 
of the two-dimensional theory along the residual circle $S^1_C$, given by an element of the flavor symmetry
of the $\mathbb{CP}^{N-1}$-model;
\begin{align}
&\mathbb{Z}_N^{(B)}: [\dots, z_{k}, \dots] \to [ \dots, e^{\frac{2\pi \ii}{k}} z_{k},  \dots, ] \;.
\end{align}

In four-dimensional language, 
this $\mathbb{Z}_N$ (zero-form) symmetry arises from the four-dimensional $\mathbb{Z}_N$ (one-form) center symmetry
by compactification along the circle $S^1_B$ (and hence the notation).\footnote{The center symmetry is a one-form symmetry \cite{Gaiotto:2014kfa},
and hence upon a circle compactification generates a zero-form symmetry
in addition to a one-form symmetry.}
Similarly, the four-dimensional $\mathbb{Z}_N$ center symmetry compactified on 
the another cycle $S^1_A$ generates another zero-form global symmetry of the 
two-dimensional symmetry. This is given by \cite{Yamazaki:2017ulc}
\begin{align}
&\mathbb{Z}_N^{(A)}: [z_1, \dots, z_{N-1}, z_N] \to [z_2, \dots, z_{N-1}, z_1] \;.
\label{ZN_A}
\end{align}
As we will see, this symmetry will play a crucial role in what follows.

\section{Twisted Compactification of 't Hooft Anomaly}
Let us now come to the 't Hooft anomalies.

As already mentioned above, the four-dimensional pure $\SU(N)$ Yang-Mills theory has the $\mathbb{Z}_N$ center one-form symmetry; we denote the corresponding
two-form discrete $\mathbb{Z}_N$ gauge field as $B$. In addition the theory has a $\mathbb{Z}_2$ 
time-reversal symmetry $\T$.

The results of \cite{Gaiotto:2017yup} shows that the pure Yang-Mills theory with 
theta angle $\theta=\pi$
has a mixed 't Hooft anomaly 
between the center symmetry and the time-reversal symmetry. This means that the 
partition of the theory $\mathcal{Z}^{\rm 4d}_{\theta=\pi}[(A,B)]$ at theta angle $\theta=\pi$, regarded as a function of the background gauge fields
$A$ and $B$, is not invariant under the time-reversal symmetry: 
\begin{align}
\mathcal{Z}^{\rm 4d}_{\theta=\pi}[\mathsf{T} \cdot (A, B)]=\mathcal{Z}^{\rm 4d}_{\theta=\pi}[(A,B)] \, \exp\left(
\frac{\ii N}{4\pi} \int B\wedge B
\right) \;.
\label{4d_anomaly}
\end{align}

Let us next compactify the theory onto the geometry $\mathbb{R}\times S_1^{(A)}\times S_1^{(B)}\times S_1^{(C)}$.
We take the limit $L_A, L_B, L_C\to 0$, while still keep the hierarchy of scales as in \eqref{hierarchy}.
By decomposing the two-form gauge field $B$ into components, we find the decomposition
\begin{align}
\begin{split}
B=
&B_A^{(1)} dx^A+B_B^{(1)} dx^B+B_C^{(1)} dx^C\\
&\quad +B_{BC}^{(0)} dx^B\wedge dx^C 
+B_{CA}^{(0)} dx^C\wedge dx^A +B_{AB}^{(0)} dx^A\wedge dx^B\;,
\label{B_decomp}
\end{split}
\end{align}
where $B_A^{(1)}, B_B^{(1)},B_C^{(1)}$ and $B_{BC}^{(0)}, B_{CA}^{(0)},B_{AB}^{(0)}$
are one-forms and zero-forms on the residual $\mathbb{R}$-direction, respectively, 
and none of them have any non-trivial dependence along the three-torus $T^3$.

The one-forms $B^{(1)}$ and the zero-forms $B^{(0)}$ play the role of the 
`electric field' and `magnetic field' for 
the discrete $\mathbb{Z}_N$ center symmetry.
The electric objects are Wilson lines (holonomies)
around the non-trivial cycles of the three-torus.
The magnetic gauge field, on the other hand, represents the 
't Hooft magnetic flux along a two-torus \cite{tHooft:1979rtg,Witten:2000nv}. For example,
the zero-form field $B_{BC}^{(0)}$
represents the Aharanov-Bohm-type phase penetrating through 
the two-torus $T_{BC}=S^1_B\times S^1_C$, making the 
holonomies along $S^1_B$ and $S^1_C$ non-commutative.

Let us substitute the decomposition \eqref{B_decomp} into the expression for the mixed anomaly \eqref{4d_anomaly}.
After trivially integrating over the small three-torus directions, 
we find that 
the mixed anomaly  \eqref{4d_anomaly} now is expressed as an integral 
over the residual $\mathbb{R}$-direction:
\begin{align}
\frac{2\ii N}{4\pi} \int_{\mathbb{R}}  
 \left(  
 B_A^{(1)} B_{BC}^{(0)} +B_B^{(1)} B_{CA}^{(0)}+B_C^{(1)} B_{AB}^{(0)}
 \right) \;.
\end{align}
The expression appearing here is an analog of the 
Poynting vector of electromagnetism, but now for 
the discrete $\mathbb{Z}_N$ center symmetry.

To this point we have not used any information regarding the 
choice of the boundary conditions.
Recall from section \ref{sec.4d2d} we turn on the 't Hooft magnetic flux is turned on along the
two-torus $T^2_{BC}=S^1_B\times S^1_C$ 
directions, and not in other directions involving $S^1_A$.
Moreover, the value of $B_{BC}^{(0)}$ can be 
derived from the fact that we have one unit of the 
't Hooft discrete magnetic flux \cite{Yamazaki:2017ulc}:
\begin{align}
\begin{split}
U_C U_B=e^{-\frac{2\pi \ii }{N}} U_B U_C \;,
\end{split}
\end{align}
where $U_B$ and $U_C$ denotes the holonomy along the $S^1_B$ and $S^1_C$.\footnote{For gauge invariance
under the twisted boundary condition a care is needed for the definition of $U_C$. Note that $U_C$ here is denoted by $U_C'$ in \cite{Yamazaki:2017ulc}.}
We thus obtain
\begin{align}
B_{BC}^{(0)}=-\frac{2\pi }{N} \;, \quad B_{CA}^{(0)}=B_{AB}^{(0)}=0 \;,
\end{align}
and we arrive at the 't Hooft anomaly
\begin{align}
\frac{2\ii N}{4\pi} \int_{\mathbb{R}}    B_A^{(1)} \left(-\frac{2\pi }{N} \right)
=-\ii \int_{\mathbb{R}}    B_A^{(1)}  \;.
\label{anomaly_2d}
\end{align}
In other words, under the time-reversal symmetry the one-dimensional partition function $\mathcal{Z}^{\rm 1d}$
at $\theta=\pi$, as a function of the background gauge field $B_A^{(1)}$,
transforms non-trivially as
\begin{align}
\mathcal{Z}^{\rm 1d}_{\theta=\pi}[\mathsf{T} \cdot B_A^{(1)}]=\mathcal{Z}^{\rm 1d}_{\theta=\pi}[B_A^{(1)}] \, \exp\left(
-\ii \int_{\mathbb{R}}    B_A^{(1)}
\right) \;.
\label{2d_anomaly}
\end{align}

It turns out that this is exactly the 't Hooft anomaly derived in \cite{Tanizaki:2017qhf}, 
which discussed the $\mathbb{Z}_N$-twisted circle compactification of the two-dimensional anomaly discussed in \cite{Komargodski:2017dmc}.
Indeed, recall that the one-form gauge field $B_A^{(1)}$
\eqref{B_decomp} arises from the reduction of the two-form field along the circle $S^1_A$,
and hence should be associated with the zero-form global symmetry $\mathbb{Z}_N^{(A)}$.
As we have seen before, this symmetry acts on the homogeneous coordinates of $\mathbb{CP}^{N-1}$ by \eqref{ZN_A}.
This is nothing but the $\mathbb{Z}_N$ `shift symmetry' (denoted by $(\mathbb{Z}_N)_S$) in \cite{Tanizaki:2017qhf},
and hence our result \eqref{anomaly_2d} coincides with (3.13) of \cite{Tanizaki:2017qhf}.
This completes our discussion of the twisted compactification of the 
't Hooft mixed anomalies.

\section{Discussion}
The result of this note proves that the vacua of the two theories, 
namely
four-dimensional pure $\SU(N)$ Yang-Mills theory on a center-twisted three-torus on the one hand, and 
two-dimensional $\mathbb{CP}^{N-1}$ model on a flavor-twisted circle on the other,
are constrained by the same 't Hooft anomaly.
In particular we find that neither theory has a trivial vacuum.

It would be interesting to extend the analysis to four-dimensional Yang-Mills theory coupled with matters,
say with adjoint or fundamental/anti-fundamental matters. 

Our results provides a rather non-trivial non-perturbative consistency check of the proposal of \cite{Yamazaki:2017ulc},
and make it even more plausible the optimistic scenario 
that the setup of \cite{Yamazaki:2017ulc} provides  a right direction towards 
an analytic demonstration of the confinement and the mass gap of the asymptotically-free 
pure Yang-Mills theory, the holy grail of the subject.

The finding of this note also supports the claim in \cite{Yamazaki:2017ulc} that 
four-dimensional pure Yang-Mills theory with theta angle has $N$ metastable vacua,\footnote{In the large 
$N$ limit, this is consistent with the old results of \cite{Witten:1980+1998} in four dimensions.}
as expected from the presence of the $N$ classical vacua in the $\CPN$-model \cite{Dunne:2012ae,Dunne:2012zk,Yamazaki:2017ulc}.
In addition to theoretical curiosity, this has an interesting implication 
to the observability of the tensor modes in the recently-proposed axion-type model of inflation \cite{Nomura-Y}.

\section*{Acknowledgements}

This work is inspired by talks and conversations at the KITP conference ``Resurgence in Gauge and String Theory''
and the KITP program ``Resurgent Asymptotics in Physics and Mathematics''
(in particular a presentation by Yuya Tanizaki),
and the author would like to thank KITP, UCSB for hospitality.
He would also like to thank Tatsuhiro Misumi, Norisuke Sakai, Yuya Tanizaki and Mithat \"Unsal
for valuable discussions.
This work is supported by WPI program (MEXT, Japan), by JSPS Program No.\ R2603, by JSPS KAKENHI Grant No.\ 15K17634, by JSPS-NRF Joint Research Project, and by NSF under Grant No.\ PHY-1125915. 


\end{document}